\documentclass{mem}
\usepackage{txfonts}
\usepackage{balance}
\usepackage{graphicx}
\usepackage[a4paper]{hyperref}
\idline{75}{281}

\begin{document}

\title{
Dark matter distribution and its impact on the evolution of galaxy disks 
}
   \subtitle{}
\author{
Fran\c{c}oise \,Combes
          }

  \offprints{F. Combes}
 
\institute{
Observatoire de Paris, LERMA, CNRS,
61 Av. de l'Observatoire, 
F-75014, Paris, France \\
\email{francoise.combes@obspm.fr}
}

\authorrunning{Combes }

\titlerunning{Dark matter and disk evolution}

\abstract{
The role of dark matter halos in galaxy disk evolution is reviewed, in
particular the stabilisation of disks through self-gravity reduction,
or the bar development through angular momentum exchange.
Triaxial dark halos tend to weaken bars.
When the dark mass inside the bar region is negligible,
the bar develops through angular momentum exchange between inner and outer disk,
and between stars and gas. Self-regulating cycles on the bar strength may develop
in the presence of external gas accretion.
Dynamical friction on dark halos slows down bars, which puts
constraints on the dark matter amount inside the inner disk.
During galaxy formation, baryons can lose most of their angular momentum
if the infall is misaligned with the dark matter axes. Stable disks can form
aligned with the minor axis of the dark halo.
 A sudden change in the infall direction, otherwise steady, can produce
the peculiar polar ring galaxies. The dark matter halo can then be aligned along
the polar disk.
Misaligned infall can also maintain lopsidedness, which is only
rarely due to galaxy interactions and mergers.

}
\maketitle{}

\section{Introduction}
 Dark matter halos are known to have a significant influence on
the stability of disks, first through a positive effect (e.g. Ostriker \& Peebles 1973),
but also a negative one, in allowing bars to develop (e.g. Athanassoula 2002). 
The radial distribution of dark matter has also a large influence, whether
the dark matter is highly concentrated in a cusp (Navarro, Frenk \& White 1997),
or flattened in a core as observed, which might be obtained through stellar feedback (e.g. Maccio et al. 2012).
 The non-dissipative dark matter component cannot share the same instabilities
as the baryonic disk, but plays the role of a reservoir in the exchange of angular 
momentum. Through dynamical friction, it can slow down bars, and the measure
of bar pattern speeds can constrain the amount of dark matter in central regions
of disk galaxies. Finally, dark matter and disk could be misaligned, according
to the formation history, and the consequences might lead to lopsidedness,
or the presence of warps and polar rings. These phenomena are now
better known thanks to detailed numerical simulations.

\begin{figure*}[ht!]
\centerline{
\resizebox{11cm}{!}{\includegraphics[angle=-0,clip=true]{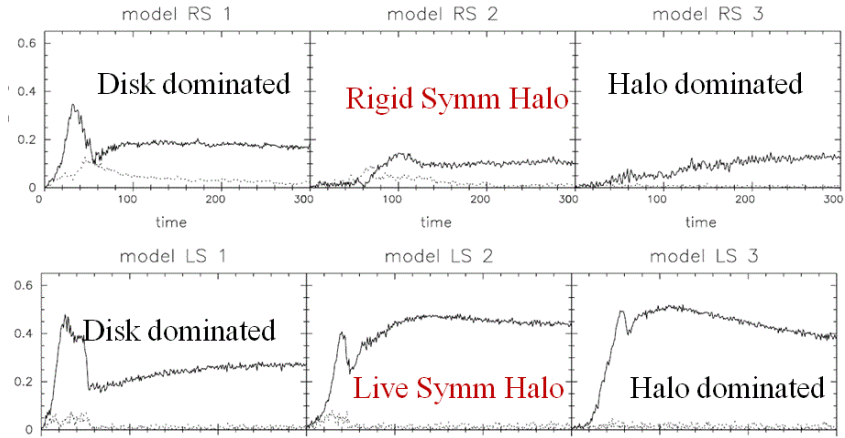}}
}
\caption{
 Numerical models of the evolution of a galaxy disk embedded
in a dark matter halo of various mass: the Fourier component
A2 (m=2, full line), and A3 (m=3, dotted line),
are plotted versus time, 
in units of 47 Myr, adapted from Berentzen et al (2006).
 From left to right, the mass ratio between halo and disk increases,
such that the model is close to maximal disk at left, and halo dominated at right.
Comparison between bar strengths with rigid halos (top) and live halos
(bottom) shows that the strongest bars are obtained with live massive halos,
in spite of the stabilizing effect of halos on disks.
}
\label{beren}
\end{figure*}

\section{Disk evolution, stability}

To quantify the influence of dark matter on disk instability, it is first interesting to 
study the dynamics of purely stellar systems, without any dark matter halos.
A series of N-body simulations have been done in the 1990's years,
showing the influence of the initial dynamical state and the gravitational heating 
in the stability of pure stellar disks, with more or less massive bulges
(see e.g. Sellwood 1987, Combes et al. 1990, Friedli \& Benz 1993).
 It is not sufficient to start the simulation with a disk in equilibrium,
with a Toomre parameter equal to Q=1-2 to ensure stability with respect to bar
growth. A bar instability will develop, after a transient spiral structure has
transfered angular momentum to the outer disk, more or less quickly according
to the initial value of Q. Some overshooting could occur, in the sense that
initially colder disks develop spirals and bars more violently, and are heated
more by gravitational instabilities, so that the final bar is weaker than in 
hotter intial disks. A more efficient way to stabilize disks with respect to
bar formation is to select initially a Toomre-parameter profile, varying
with radius, taking higher values in the center (e.g. Athanassoula \& Sellwood 1986).

About the bar pattern speed, its evolution has also been followed during
bar growth in only baryonic galaxies, with stars and gas: the bar starts as a fast rotator,
and then continuoulsy slows down, to stabilize after
one or two billion years (Combes et al. 1990, Friedli \& Benz 1993).
This slowing down is due to more and more orbits trapped into the bar,
and to the orbits becoming progressively more elongated: the precessing rates
of these orbits is lower when their elongation is larger.
Secular evolution occurs with little dark matter halo through 
angular momentum exchange between gas and stars, and from the inner 
to outer parts.  
The exchange with the gas component may introduce a cycle, provided that 
external gas accretion is considered (Bournaud \& Combes 2002,
Bournaud et al 2005a). The bar torques on the gas remove its angular momentum,
and drives gas inflow. The gas gives its angular momentum to the bar, which 
weakens or disappears. Gas accretion is then able to replenish the gas, and trigger
another bar instability in the cooled disk.

\section{Angular momentum transfers, bar evolution}

After the pioneering work of Ostriker \& Peebles (1973) and for a long time,
it was thought that the only influence of a dark matter halo was to stabilize disks,
in lowering the critical velocity dispersion to reach a Toomre parameter Q$\sim$ 1.
Indeed,  the critical dispersion is proportional to the disk surface density,
and inversely proportional to the epicyclic frequency, and the latter is largely
increased by the presence of the dark matter halo. This is indeed what is found
when the halo component is rigid (e. g. Figure \ref{beren}).
 However, when the halo is live, and able to exchange angular momentum
with the baryons in the disk, it favors the formation of a bar (Athanassoula 2002).
  It is then possible to form stronger bars when the halo is more massive,
as shown in  Figure \ref{beren}. However, when the dark matter dominates
the mass inside the bar, another phenomenon occurs: dynamical friction of
the tumbling bar against the particle of the dark matter halo, slowing down the bars,
as will be discussed in next section.

The dark matter halo could also be triaxial, as frequently found in cosmological
simulations. The triaxiality of halos then destroy bars (Berentzen et al 2006). Indeed, the bar is also a triaxial
structure, but misaligned, and with different pattern speed than the triaxial halo. There is 
no compatible resonances, and the existence of these two triaxial structures
generates chaos in the stellar orbits, weakening the bar. This is different from the situation of
embedded bars, where the nuclear bar rotates much faster than the primary, and it is possible to have
a common resonance. Then embedded bars could survive for several rotations.

\begin{figure}[ht!]
\centerline{
\resizebox{7cm}{!}{\includegraphics[angle=-0,clip=true]{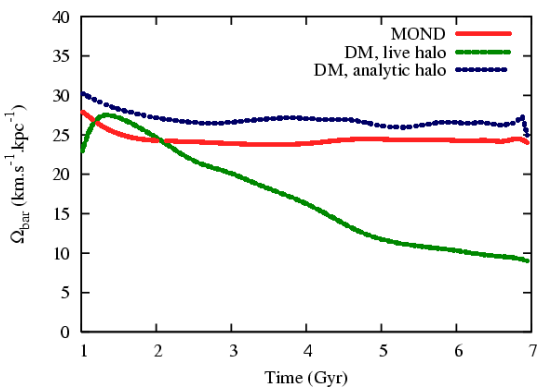}}
}
\caption{
Evolution of the pattern speed of the bar in several models. In the case of newtonian gravity with 
dark matter halos, the pattern speed remains constant when the halo is maintained rigid,
but the bar slows down when the halo is live. With the modified gravity (MOND), where the
same rotation curve is accounted for without any dark matter halo, 
the bar rotates with the same velocity from the 
beginning to the end (from Tiret \& Combes 2007).
}
\label{omega}
\end{figure}
\section{Bar pattern speeds}

The dynamical friction of bars in dark matter halos was computed analytically by Weinberg (1985),
who concluded that bars would be slowed down in a few rotations. Hernquist \& Weinberg (1992)
confirmed with simulations a very short time-scale of less than a billion yr, considering a rigid bar.
 With fully consistent simulations without gas, Debattista \& Sellwood (1998) showed that stellar
bars in a dark matter halo indeed slow down very quickly, and this can put strong
constraints on the amount of dark matter present within the bar radius, since the observations
are favoring fast bars, ending at their corotation (Debattista \& Sellwood 2000).
The interaction between the stellar disk and the dark matter halo occurs essentially
at resonances (Athanassoula 2003), and the result is to increase the rotation
of the halo, which could also reveal some kind of a bar instability.
The exchange of angular momentum between the bar and the live halo is
inevitable in dark matter embedded disks, even in maximal disks models. As shown in 
Figure \ref{omega}, the slowing down of the bar pattern speed is stopped for simulations  carried out
in the frame of the MOND modified gravity. The bar then develops through 
angular momentum exchange between inner and outer disk.

\begin{figure*}[ht!]
\centerline{
\resizebox{11cm}{!}{\includegraphics[angle=-0,clip=true]{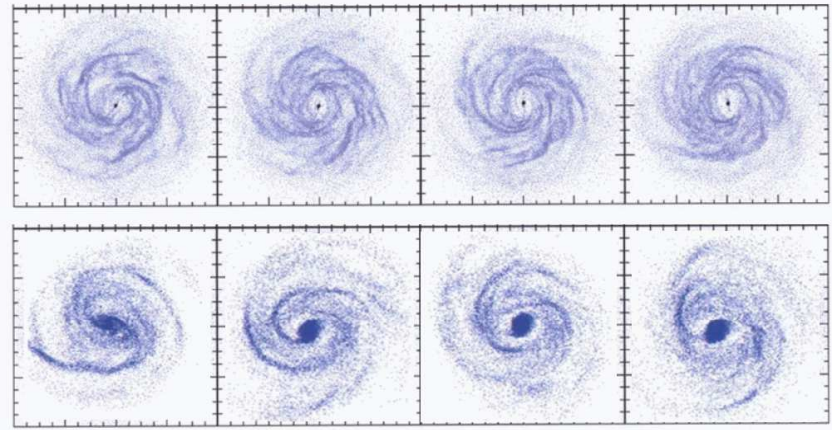}}
}
\caption{
Influence of the radial distribution of dark matter on disk evolution.
The top simulation (four snapshots every 200 Myr) is made with a cored dark matter of the same
total mass inside the optical disk than the bottom simulation with
an NFW profile. Much more structure develops towards the center in
the top simulation, the gravity torques produce gas flows and
a central concentration; while in the bottom simulation, the dominant
dark matter concentration prevents the non-axisymmetric structures and 
the central gas flows to develop.
}
\label{nfw-core}
\end{figure*}
\section {Disk formation in dark matter halos}

The presence of a dense concentration of dark matter has some influence in the secular 
evolution through bar and spiral waves, in the sense that it forces an axisymmetric potential in 
the center, and weakens the gravitational effect of the bar on the gas.
 Figure \ref{nfw-core} shows how the gas inflow due to bar torques is suppressed when 
a cuspy dark matter distribution is present. The absence of non-axisymmetric features
in the central parts slows down the concentration of the baryonic disk.

Galaxies are supposed to form when baryons fall into dark matter halos,
which could be somewhat misaligned with the main orientation of the baryons. 
 Continuous infall of material from the inter-galactic medium makes
the galaxy grow, although it may change orientation regularly, due to
slight misalignment. The torques and angular momentum exchange have been recently
studied by Aumer \& White (2012).  Dark matter halos are extracted from a large
cosmological simulation (Aquarius, Springel et al 2008), and  resimulated with 700 pc resolution.
Then baryons are launched at z=1.3 in the dark halos, as rotating spheres of hot gas (at 10$^6$ K).
Progressively the gas cools and forms stars. The dissipation of the baryons leads to an oblate
system, which progressively modifies the triaxial dark matter potential, transforming 
into an axisymmetric shape. These particular simulations ignore cold streams,
and use a temperature floor of 10$^5$K to prevent clumping of the gas disk into
fragments. The disk forms inside-out, with a break in surface density in the outer parts,
with formation of a warp, due to late infall. The disk distribution is exponential,
with a large central concentration, implying a large angular momentum loss.
The phase of angular momentum loss is simultaneous to the axi-symmetrisation of the
dark halo potential. Extreme losses of momentum occur when the infall of
baryonic matter is misaligned with the dark halo axes.
 Stable disks tend to align with the minor axis of the halo. If the infall 
is from the start towards the major axis, no settlement is possible, and 
the system contracts to a compact object, having lost most of its angular momentum.
A conclusion could be that disks form when the baryonic infall happens to be
aligned with the halo axes, and on the contrary, spheroids will form
in case of misalignment. 

\section {Polar rings, lopsidedness}

In the special case of almost orthogonal misalignment, peculiar objects with two 
perpendicular disks could form, similar to polar ring galaxies,
as simulated by Brook et al (2008): the most likely scenario
is a sudden change by 90$^\circ$ in the direction of baryonic infall and accretion.
The initial infall direction accounts for the primary galaxy system, esuatorial, and 
the second direction gives rise to the polar system. 
 This peculiar formation scenario is important to understand the shape of dark matter halos
observed in polar ring galaxies (PRG). The kinematics of stars and gas have been studied
intensively in the main known PRGs in order to derive the 3D shape of dark matter halos,
but surprising results were found, the dark matter being aligned with the polar systems 
(e.g. Iodice et al 2003). The polar component in general is quite massive in PRGs
(due to selection effects) and the baryons settled in this disk cannot be taken as 
test particles to probe the potential. Snaith et al (2012) analysis of the simulations
confirms that the polar system is similar to a new disk formed after a last merging
event, and subsequent gas infall. The persistence of the two orthogonal systems
comes from the fact that the infall direction is coherent and stable during 
a few billion years.

The misalignment of baryonic infall and dark matter halo axes has other consequences,
such as triggering m=1 instabilities, or lopsidedness in galaxy disks (e.g. Jog \& Combes 2009).
Jog (1997, 1999) showed that the response of a galaxy disk to a lopsided halo is important
mainly at large radii, and this can easily explain the
lopsidedness in the atomic hydrogen gas observed in the outer disks. 
Bournaud et al (2005b) proposed several scenarios  to explain the origin of the observed disk lopsidedness.
 A first obvious origin could be tidal interactions. However, they are not frequent enough,
and statistically the main lopsided disks are observed in isolation. It is possible that 
these isolated galaxies had a recent minor merger in the past. But simulations show
that the lopsided perturbations due to a minor merger are short-lived, and the frequency
of minor mergers is not sufficient to account for the observed statistics of lopsidedness.
The favored scenario is to rely on external gas accretion, which is intermittent but
with sufficient frequency. Accretion is not isotropic at a given time, but follows the cosmic
filaments. It can trigger easily lopsidedness at large scale. The persistence of
the lopsided morphology, given the rate of cosmological accretion, is compatible
with observations. The scenario also explains why late-type galaxies are found to be more lopsided, 
and why m=2 spiral arms and bars are correlated with disk lopsidedness.

\section{Conclusion}

Dark matter halo distribution and shape have large influence on
the formation and evolution of galaxy disks. In particular:

(1) The dark halos reduce the self-gravity of disks and stabilise them
against gravitational instabilities and namely againt bars.
When the dark halo is simulated live and axi-symmetric, it accepts angular
momentum from the stellar disk, and can favor on the contrary the 
formation of strong bars. Triaxial halos however generate chaos in a barred galaxy,
and weaken the bar instability.

(2) When the disk inside the bar region is not dominated by dark matter,
then the bar develops through angular momentum exchange between inner
and outer disk, and between stars and gas. A self-regulated cycle of bar
growth and weakening can develop in the presence of external gas accretion.
Bar torques drive gas inflow, the gas providing its angular momentum to the bar,
weakening the bar.

(3) Dark matter halos through dynamical friction can slow down bars
very efficiently, when they are dominating the mass inside the bar region. This puts 
constraints on the amount of dark matter to explain the observation of fast bars. 

(4) Dark matter halos remove angular momentum from baryons, during
galaxy formation, especially when the baryonic infall is misaligned with 
the minor axis of the halo. The latter is the configuration providing
the most stable disks. In case of misalignment, the loss of angular momentum
is so large than only compact spheroids form. When the accretion 
is steady during Gyrs, but change suddenly by $\sim$ 90$^\circ$, two orthogonal systems
may form a polar ring.  In some cases, the dark matter may appear aligned
with the polar disk.

(5) Misalignment between baryons and halo can also trigger lopsidedness, 
in particular in the outer disks.  The observed statistics of lopsided stellar disks
cannot be easily explained through galaxy interactions nor minor mergers,
but must rely in a large part on external gas accretion from cosmic filaments.

\begin{acknowledgements}
Many thanks to the organisers of this exciting Special Session,
Paola Si Matteo and Chanda Jog.
\end{acknowledgements}

\bibliographystyle{aa}

\begin{thebibliography}{}
\bibitem{} Athanassoula E.: 2003, MNRAS 341, 1179
\bibitem{} Athanassoula E.: 2002, ApJ 569, L83
\bibitem{} Athanassoula E., Sellwood J.: 1986 MNRAS 221, 213
\bibitem{} Aumer M., White S.D.M.: 2012, MNRAS in press (arXiv1203.1190)
\bibitem{} Berentzen, I., Shlosman, I., Jogee, S.: 2006 ApJ 637, 582
\bibitem{} Bournaud, F., Combes, F., Semelin, B.: 2005a, MNRAS 364, L18
\bibitem{} Bournaud, F., Combes, F., Jog C.J., Puerari I.: 2005b, A\&A 438, 507
\bibitem{} Bournaud, F., Combes, F.: 2002, A\&A
\bibitem{} Brook, C. B., Governato, F., Quinn, T. et al.: 2008, ApJ 689, 678 
\bibitem{} Combes F., Debbasch F., Friedli D., Pfenniger D.: 1990, A\&A 233, 82
\bibitem{} Debattista V., Sellwood J.:  2000, ApJ 543, 704
\bibitem{} Debattista V., Sellwood J.:  1998, ApJ 493, L5
\bibitem{} Friedli D., Benz W.: 1993, A\&A 268, 65
\bibitem{} Hernquist L., Weinberg M.D.: 1992, ApJ 400, 80
\bibitem{} Iodice E., Arnaboldi, M., Bournaud, F. et al.: 2003, ApJ 585, 730 
\bibitem{} Jog C.J., Combes F.: 2009, Physics Reports, 471, 75
\bibitem{} Jog C.J.: 1999 ApJ 522, 661
\bibitem{} Jog C.J.: 1997 ApJ 488, 642
\bibitem{} Maccio A. V., Stinson, G., Brook, C. B. et al. 2012 ApJ 744, L9
\bibitem{} Navarro, J. F., Frenk, C. S., White, S. D. M. 1997 ApJ 490, 493
\bibitem{} Ostriker, J. P., Peebles, P. J. E. 1973 ApJ 186, 467
\bibitem{} Sellwood J.: 1987 ARA\&A 25, 151
\bibitem{} Snaith, O. N., Gibson, B. K., Brook, C. B. et al. 2012, MNRAS 425, 1967 
\bibitem{} Springel, V., Wang, J., Vogelsberger, M. et al.: 2008, MNRAS 391, 1685 
\bibitem{} Tiret O., Combes F.: 2007, A\&A 464, 517
\bibitem{} Weinberg M.D.: 1985, MNRAS 213, 451

\end{thebibliography}

\end{document}